\documentclass[twocolumn]{article}
\usepackage{amssymb}
\usepackage{german} 
\begin{document}
\tolerance=10000
\def\shalf{\hbox{${\textstyle{\frac{1}{2}}}$}}

\title{"`...ich dachte mir nicht viel dabei..."'\\ 
       Plancks ungerader Weg zur Strahlungsformel}

\author{Domenico Giulini und  Norbert Straumann \\
        Institut f"ur Theoretische Physik       \\
        der Universit"at Z"urich                \\
        Winterthurerstrasse 190                 \\
        CH-8057 Z"urich, Schweiz                  }
  
\date{Oktober 2000}
\maketitle

\subsection*{Zusammenfassung}
\textbf{Max Plancks Ableitung seiner Strahlungsformel,
mitgeteilt am 14.~Dezember~1900 und von ihm selbst als 
"`Akt der Verzweiflung"' charakterisiert, markiert
die Geburtsstunde der Quantentheorie. Gleichzeitig wurde 
Planck dadurch zur Aufgabe eines langj"ahrigen und systematisch 
angelegten Forschungsprogramms gezwungen, in dem er mit Hilfe 
der gerade erst etablierten Maxwell'schen Elektrodynamik versuchte, 
den 2.~Hauptsatz der Thermodynamik als streng deterministisches 
Gesetz zu begr"unden.
}

\subsection*{Einleitung}
Die Studienzeit Max Plancks (1858-1947) f"allt in die "`besten 
Mannesjahre"' der klassischen theoretischen Physik, die in 
Deutschland durch das stattliche Dreigestirn der Mittf"unfziger 
Clausius, Helmholtz und Kirchhoff eindr"ucklich vertreten war.
Von diesen haben vor allem die Schriften Clausius' den Studenten 
Planck durch ihre \emph{"`Klarheit und "Uberzeugungskraft der Sprache"'} 
besonders angezogen. Mit seiner bis heute unver"andert gelehrten 
Formulierung des 2.~Hauptsatzes der Thermodynamik von 1865 gab  
Clausius ein Musterbeispiel des Planck'schen Ideals einer 
physikalischen Gesetzm"a"sigkeit:
\emph{"`Was mich in der Physik von jeher vor allem interessierte, 
waren die gro"sen allgemeinen Gesetze, die f"ur s"amtliche 
Naturvorg"ange Bedeutung besitzen, unabh"angig von den 
Eigenschaften der an den Vorg"angen beteiligten K"orper."'} 
\footnote{Dieses und das vorangegangene Zitat Plancks sind 
seinem Aufsatz "`Zur Geschichte der Auffindung des physikalischen 
Wirkungsquantums"' entnommen; \cite{Planck-GW}, Bd.3, p.255.}
So ist es nicht verwunderlich, dass sich Planck eben diesen
2.~Hauptsatz auch zum Thema seiner Dissertation w"ahlte.

Neben der Thermodynamik beherrschte vor allem die Elektrodynamik 
die physikalische "Offentlichkeit im letzten Viertel des
19. Jahrhunderts. Der feldtheoretischen Formulierung
Maxwells standen in Deutschland die Fernwirkungstheorien 
Webers und deren leichte Modifikation durch Helmholtz entgegen.
Erst Heinrich Hertz, nur ein Jahr "alter als Planck, verhalf 
ersterer sowohl mit seinen Aufsehen erregenden Versuchen zum
Nachweis elektromagnetischer Wellen als auch mit seinen 
theoretischen Untersuchungen endg"ultig zum 
Durchbruch (siehe \cite{Foelsing-Hertz}). Somit zerfiel  
der physikalische Materiebegriff in die Dualit"at 
von Materie und (elektromagnetischem) Feld. Letzteres
wurde damals noch als Anregungszustand des hypothetischen 
Mediums \emph{"Ather} aufgefasst, der als schwerelos galt und 
der ponderablen (d.h. w"agbaren) Materie gleichberechtigt an die 
Seite gestellt wurde. Einen hervorragenden Eindruck dieser 
Auffassung vermittelt die gerade erstmalig herausgegebene 
Hertz'sche Vorlesung "uber \emph{die Constitution 
der Materie} \cite{Hertz-CDM} aus dem Jahre 1884.
Mit dem Durchsetzen der Speziellen Relativit"atstheorie von 1905 
ist dann bekanntlich auch die "Athervorstellung einem abstrakten, 
nicht substantiellen Feldbegriff gewichen. 

Durch dieses dualistische Materiekonzept entstand unweigerlich 
die Frage nach der Natur der Wechselwirkung von Strahlung mit 
Materie, insbesondere nach den Mechanismen der Erzeugung und 
Vernichtung von Strahlung. 
Modellm"a"sig gelang zuerst Heinrich Hertz 1888 die vollst"andige 
Beschreibung der Emission elektromagnetischer Strahlung durch 
strenges L"osen der Maxwell'schen Gleichungen f"ur den Spezialfall 
eines harmonisch schwingenden elektrischen Dipols (seitdem `Hertz'scher
Oszillator' genannt). Was hingegen die \emph{allgemeinen}, von 
idealisierenden Modellvorstellungen unabh"angigen Gesetzm"a"sigkeiten
anbelangt, zeigte sich einmal mehr die ungeheure Kraft 
thermodynamischer "Uberlegungen. Dabei bestand nach den
Hertz'schen Entdeckungen kein Zweifel mehr, dass W"armestrahlung dem 
Licht wesensgleich und nur durch die Wellenl"ange unterschieden ist
und somit durch die Maxwell'sche Theorie im Prinzip vollst"andig
beschrieben werden kann. Man durfte somit einfach von "`Strahlung"' 
reden.

\subsection*{Fr"uhe Strahlungstheorie}
Thermodynamische "Uberlegungen auf der Basis des 
2.~Hauptsatzes f"uhrten Kirchhoff schon 1859 zu der Einsicht, dass 
in einem gleichtemperierten Raum, dessen W"ande f"ur Strahlung 
undurchl"assig sind, die sich einstellende Gleichgewichtsstrahlung 
von der Form des Raumes und der Natur der in ihm enthaltenen 
K"orper unabh"angig ist. Diese Strahlung ist identisch derjenigen, 
welche ein vollkommen schwarzer K"orper aussenden w"urde. Dabei f"uhrte 
Kirchhoff die Idealisierung des \emph{schwarzen K"orpers} ein,
der dadurch definiert ist, dass sein Absorptionsverm"ogen den 
h"ochsten theoretisch erreichbaren Wert 1 besitzt.
Es muss also eine universelle Funktion $u(T,\nu)$ geben,
die die spektrale Energiedichte der Strahlung im Gleichgewicht 
bei der absoluten Temperatur $T$ und im Frequenzintervall 
$[\nu,\nu+d\nu]$ angibt. Die Aufgabe war nun, diese Funktion 
zweier unabh"angiger Variablen zu bestimmen.
F"ur eine konzise Darstellung der fr"uhen Strahlungstheorie 
sei auf \cite{Lorentz-1927} verwiesen.

Zun"achst wurden weitere Informationen "uber die Funktion 
$u(T,\nu)$ gewonnen. So folgt aus der Maxwell'schen Theorie, 
dass ein isotropes Strahlungsfeld auf die undurchl"assige Bewandung 
einen Druck aus"ubt, der gleich $1/3$ der gesamten, d.h. "uber 
alle Frequenzen integrierten Energiedichte $U(T)$ ist. 
Durch eine einfache thermodynamische Betrachtung konnte  
Boltzmann 1884 damit das von Stefan empirisch ermittelte 
\emph{Stefan-Boltzmann'sche-Gesetz} ableiten:
\begin{equation}
U(T):=\int_0^{\infty}u(T,\nu)\,d\nu=\sigma T^4\,,
\label{Stefan-Boltzmann}
\end{equation}
wobei $\sigma$ eine Konstante ist.

Einen wesentlichen Fortschritt brachte eine Arbeit von W.~Wien
aus dem Jahre 1893, in der er durch recht raffinierte thermodynamische 
"Uberlegungen bewies, dass $u(T,\nu)$ von folgender Form sein muss
\emph{(Wien'sches Verschiebungsgesetz)}:
\begin{equation}
u(T,\nu)=\nu^3\,f(\nu/T)\,.
\label{Wien}
\end{equation}
Damit war das "`Strahlungsproblem"' auf die Bestimmung der einen 
universellen Funktion $f$ \emph{einer} Ver"anderlichen zur"uckgef"uhrt.

Zu diesem Zeitpunkt gab es keinerlei Anhaltspunkte daf"ur, dass sich 
diese "uberschaubar anmutende Aufgabe zu einem der tiefgr"undigsten 
und ausdauerndsten Probleme der gesamten Physikgeschichte 
auswachsen w"urde.  Noch zwanzig Jahre sp"ater, also im Jahre 1913,  
gab Einstein in seiner Rede anl"a"slich der "Ubergabe des Rektorats der 
Berliner Universit"at an Max Planck die folgende Einsch"atzung: 
\emph{"`Es w"are erhebend, wenn wir die Gehirnsubstanz auf eine    
Waage legen k"onnten, die von den theoretischen Physikern auf dem 
Altar dieser universellen Funktion $f$ hingeopfert wurde; und es 
ist dieses grausamen Opfers kein Ende abzusehen! Noch mehr: auch die 
klassische Mechanik fiel ihr zum Opfer, und es ist nicht abzusehen, ob 
Maxwells Gleichungen der Elektrodynamik die Krisis "uberdauern werden, 
welche diese Funktion $f$ mit sich gebracht hat."'}
  
Doch zur"uck zum Wien'schen Verschiebungsgesetz (\ref{Wien}). Es
beinhaltet das Stefan-Boltzmann'sche-Gesetz (\ref{Stefan-Boltzmann}), wie 
man sofort durch Integration "uber $\nu$ einsieht. Durch 
Differentiation findet man, dass sich die Frequenz maximaler 
Energiedichte proportional zu $T$ verschiebt (woraus der Name 
\emph{Verschiebungs}gesetz resultiert), was ebenfalls ein 
damals empirisch bekanntes Gesetz war. In Anlehnung an die 
exponentielle Maxwell'sche Geschwindigkeitsverteilung machte 
Wien sogar folgenden konkreten Vorschlag (in der Schreibweise 
Plancks; $a$ und $b$ sind Konstanten) 
\begin{equation}
u(T,\nu)=\frac{8\pi\nu^3}{c^3}b\,\exp(-\frac{a\nu}{T})\,.
\label{Wien-Gesetz}
\end{equation}
Dieses \emph{Wien'sche Strahlungsgesetz} war bis etwa Mitte 
des Jahres 1900 mit den experimentellen Daten vereinbar.

\subsection*{Plancks Forschungsprogramm}
Nach dem, was wir bereits "uber Plancks wissenschaftliche
Orientierung geh"ort haben, ist es kaum verwunderlich, dass das
Problem der Bestimmung der universellen Funktion $f$ 
genau nach seinem Geschmack war. Er ging dieses Problem 
jedoch nicht direkt an, sondern bettete es ein in ein sorgf"altig 
geplantes und systematisch vorangetriebenes Forschungsprogramm, 
dessen Ziel eine strenge Ableitung des 2.~Hauptsatzes der 
Thermodynamik mit Hilfe der Maxwell'schen Gesetze
der Elektrodynamik war. Dabei ist es wesentlich, daran zu erinnern, 
dass Planck zu dieser Zeit den 2.~Hauptsatz als streng deterministisches 
Gesetz verstand und nicht in seiner wahrscheinlichkeitstheoretischen 
Bedeutung, wie wir es heute tun. "Uberhaupt hielt Planck damals noch 
wenig von Wahrscheinlickeitsgesetzen in der Physik, weil die von 
diesen stets erlaubten Ausnahmen keinen Platz in den ausnahmslos 
g"ultigen Gesetzm"a"sigkeiten hatten, mit denen er die Physik ausgestattet 
wissen wollte. Dabei war es Planck klar, dass als Folge des 
Poincar{\'e}'schen Wiederkehrsatzes ein solcher Beweis 
innerhalb der Mechanik \emph{nicht} zu f"uhren war.  So hatte etwa 
1896 sein damaliger Assistent
E.~Zermelo\footnote{E.~Zermelo hat 1908 als erster eine
Axiomatisierung der Mengenlehre vorgeschlagen, wodurch er ein bis
heute sehr bekannter Mathematiker wurde.}
einen noch heute sehr lesbaren und eleganten Beweis dieses Satzes
ver"offentlicht.

Innerhalb dieses Forschungsprogramms entstanden in kurzer Zeit 
Vorarbeiten, die sich --~an Hertz anschlie"send~-- mit der 
Theorie der 
Emission und Absorption elektromagnetischer Wellen durch einfache 
Oszillatoren besch"aftigten. Dabei entwickelte Planck "ubrigens 
ein Jahr vor Larmor (1896) die heute nach letzterem benannten Formeln
f"ur die Abstrahlungsleistung und Strahlungsd"ampfung. 
Knapp zusammengefasst war Plancks Grundgedanke folgender: 
Nach Kirchhoff stellt sich im thermodynamischen Gleichgewicht ein 
von der Beschaffenheit der Materie unabh"angiges Energiespektrum   
ein. Also ist es zul"assig, zu dessen theoretischer Berechnung die
Materie so zu idealisieren, dass sie der genauen Berechnung von 
Prozessen der Emission und Absorption zug"anglich wird. 
Dabei ist es ganz unerheblich, ob solche idealisierten 
Oszillatoren in der Natur "uberhaupt realisiert werden; es kommt 
lediglich darauf an, dass sie mit den Naturgesetzen in 
Einklang stehen.

In seiner letzten von f"unf Arbeiten \emph{"Uber irreversible 
Strahlungsvorg"ange} aus dem Jahre 1899 leitet Planck folgende,
f"ur sein weiteres Vorgehen fundamentale Beziehung zwischen 
$u(T,\nu)$ und dem zeitlichen Mittelwert $E(T,\nu)$ 
der Energie eines im Strahlungsfeld eingebetteten geladenen 
harmonischen Oszillators der Eigenfrequenz $\nu$ ab:
\begin{equation}
u(T,\nu)=\frac{8\pi\nu^2}{c^3}\,E(T,\nu)\,.
\label{Planck-Fundamentalgleichung}
\end{equation}
Die N"utzlichkeit dieser Gleichung liegt in der bemerkenswerten 
Tatsache begr"undet, dass sie die Parameter Ladung, Masse und
D"ampfungskonstante des Oszillators nicht enth"alt. 
F"ur das Folgende ist es wichtig 
zu betonen, dass sie eine unzweideutige Folge der klassischen 
(d.h. Maxwell'schen) Elektrodynamik ist. H"atte sich Planck damals 
mit der statistischen Mechanik angefreundet, so h"atte er aus deren 
"Aquipartitionstheorem als unausweichliche Folge der klassischen 
Mechanik die Formel $E=(R/N)T$ f"ur die mittlere Energie des linearen 
harmonischen Oszillators erhalten. Dabei ist $R$ die universelle, 
auf ein Mol bezogene Gaskonstante und $N$ die Avogadro-Zahl
(Anzahl der Molek"ule in einem Mol).
Damit w"are er durch (\ref{Planck-Fundamentalgleichung}) ein Jahr 
fr"uher als Rayleigh zur \emph{Rayleigh-Jeans'schen Strahlungsformel} 
gekommen 
\begin{equation}
u(T,\nu)=\frac{8\pi\nu^2}{c^3}\frac{R}{N}T\,.
\label{Rayleigh-Jeans}
\end{equation}
Nun ist aber trotz der scheinbar gut begr"undeten Herleitung 
diese  Formel v"ollig unakzeptabel: Zun"achst impliziert eine
lineare
Abh"angigkeit von $T$, dass etwa die Strahlungsdichte bei 
Raumtemperatur ($T\approx 290$K) noch einem Sechstel der Weissglut 
schmelzenden Stahls ($T\approx 1700$K) entsprechen muss, was
offensichtlich absurd ist. Weiter 
impliziert die  quadratische Abh"angigkeit von $\nu$ eine   
physikalisch sinnlose Divergenz ("`Ultraviolett-Katastrophe"')
des Integrals (\ref{Stefan-Boltzmann}) und damit der Energiedichte 
$U(T)$. Damit wird von (\ref{Rayleigh-Jeans}), d.h von der 
klassischen Physik, die Existenz eines Gleichgewichts negiert. 
Es ist oft dar"uber ger"atselt worden, warum Planck diesen
schlagenden Hinweis darauf, dass mindestens eine der beiden 
klassischen Theorien -- Mechanik und Elektrodynamik -- nicht 
uneingeschr"ankt richtig sein kann, nicht oder zumindest erst
sehr sp"at aufnahm; und dies, obwohl ab 1905  Einstein keine 
Gelegenheit auslie"s, diesen Umstand immer wieder zu betonen.

Neben diesem rein elektrodynamischen Teil besch"aftigte Planck sich 
vor allem mit der Frage nach m"oglichen Ausdr"ucken f"ur die Entropie 
der Strahlung, die er ebenfalls auf die Frage nach der Entropie des 
einzelnen Oszillators zur"uckf"uhrte. Obwohl das eigentliche Ziel 
war, mit Hilfe der Maxwell'schen Gleichungen ein strenges Anwachsen der 
Entropie in der Zeit zu 
demonstrieren\footnote{Ganz wesentlich f"ur 
das teilweise Gelingen war Plancks "`Hypothese der nat"urlichen 
Strahlung"', die der Annahme einer vollst"andigen Inkoh"arenz der 
einzelnen Strahlungsanteile entsprach. Planck sah lange nicht, dass 
wegen der Bewegungsumkehrinvarianz der Maxwellgleichungen --~auf die
er durch Boltzmann deutlich hingewiesen wurde~--, diese Annahme
absolut 
essentiell ist, analog der Annahme der "`molekularen Unordnung"'
beim Boltzmann'schen Beweis des $H$-Theorems.},
ergab sich daraus auch Plancks Strategie zur L"osung 
des Strahlungsproblems: Man bestimme die Entropie $S(E,\nu)$ des 
einzelnen Oszillators (Eigenfrequenz $\nu$) im Strahlungsfeld als 
Funktion seiner Energie $E$. Aus der thermodynamischen Relation 
$dS/dE=1/T$ erh"alt man dann durch Aufl"osen nach $E$ die Funktion 
$E(T,\nu)$ und damit aus (\ref{Planck-Fundamentalgleichung}) $u(T,\nu)$.  
Dabei reduziert (\ref{Wien}) das Problem wiederum auf das Auffinden 
einer  Funktion \emph{einer} Variablen, denn  
mit (\ref{Planck-Fundamentalgleichung}) ist (\ref{Wien}) "aquivalent 
zu $E(T,\nu)=\nu f(\nu/T)$ und damit auch zu
\begin{equation}
S(E,\nu)=f(E/\nu)\,,
\label{Wien-Entropie}
\end{equation}
wobei $f$ f"ur eine jeweils andere, noch unbekannte Funktion steht.

In der letzten seiner Arbeiten \emph{"Uber irreversible 
Strahlungsvorg"ange} (\cite{Planck-GW}, Bd.1) leitet Planck dann 
das Wien'sche Strahlungsgesetz in der eben beschriebenen Weise ab, 
wobei er allerdings die dazu n"otige Funktion 
$S(T,\nu)$ definitorisch einf"uhrt und bemerkt, dass er keinen anderen 
mit dem 2.~Hauptsatz vertr"aglichen Ausdruck hat finden k"onnen. Er
glaubt \emph{"`hieraus schlie"sen zu m"ussen, dass die gegebene 
Definition der Strahlungsentropie und damit auch das Wien'sche 
Energieverteilungsgesetz eine notwendige Folge der Anwendung des
Principes der Vermehrung der Entropie auf die elektromagnetische
Strahlungstheorie ist und dass daher die Grenzen der G"ultigkeit 
dieses Gesetzes, falls solche "uberhaupt existiren, mit denen des 
zweiten Hauptsatzes der W"armetheorie zusammenfallen."'}
So fest glaubt er zu dieser Zeit (November 1899) in Besitz einer
absoluten Wahrheit zu sein, dass er diese Arbeit mit der Einf"uhrung 
eines "`nat"urlichen Einheitensystems"' beendet, was dadurch definiert 
ist, dass in ihm die Konstanten $a$, $b$ des Wien'schen 
Strahlungsgesetzes, die Lichtgeschwindigkeit $c$ und die Newton'sche 
Gravitationskonstante $G$ jeweils den Wert eins haben. Diese heute 
mit Hilfe der Planck'schen Konstanten $h$ und der Boltzmann'schen 
Konstanten $k$ definierten \emph{Planck'schen Einheiten} stammen 
also aus einer Zeit, in der das Planck'sche Strahlungsgesetz noch 
gar nicht existierte. 

Bereits am 19. Sept. 1899 auf der Naturforscherversammlung 
in M"unchen und wiederholt in der Sitzung der DPG vom 3.~Nov. 
desselben Jahres, hatten O.~Lummer und E.~Pringsheim "uber 
systematische Abweichungen vom Wien'schen Strahlungsgesetz im 
langwelligen Spektralbereich ($4-8.5\,\mu$) berichtet.
Dadurch etwas verunsichert versuchte Planck im  M"arz 1900 erneut 
seine Begr"undung der Wien'schen Strahlungsformel durch eine 
\emph{Ableitung} der Entropiefunktion $S(T,\nu)$ aus allgemeinen 
Prinzipien zu st"arken, was ihm auch mit Hilfe einer plausibel  
scheinenden Hypothese gelang, die sich erst sp"ater als falsch
herausstellen sollte. Noch hoffte er, dass sich die von Lummer
und Pringsheim gefundenen "`Divergenzen von erheblicher
Natur"' (Planck) nicht best"atigen w"urden. 

\subsection*{Das Planck'sche Strahlungsgesetz}
Weitere Messungen an der physikalisch-technischen
Reichsanstalt durch Kurlbaum und Rubens, und ebenso durch
Paschen in Hannover, best"atigten die von Lummer-Pringsheim 
gefundenen systematischen Abweichungen, wor"uber
Kurlbaum in der Sitzung der Deutschen Physikalischen Gesellschaft 
vom  19.~Oktober 1900 vortrug. Planck, dem diese Ergebnisse 
schon vorher mitgeteilt wurden, musste damit jede Hoffnung
auf eine Ableitung des Wien'schen Strahlungsgesetzes aufgeben. 
Auch sein Hauptziel, die strenge Begr"undung des 2.~Hauptsatzes, 
schien nun in weite Ferne ger"uckt. 

Trotzdem nicht m"ude, schlug er in der gleichen Sitzung eine 
"`Verbesserung"' der Wien'\-schen Formel vor, die er gem"a"s 
seiner alten Strategie "uber $S(E,\nu)$ zu bestimmen
suchte. In seiner vorherigen Begr"undung der Wien'\-schen Formel hatte 
der 2.~Differentialquotient von $S$ nach $E$ eine zentrale Rolle 
gespielt, der im Wien'\-schen Fall gerade proportional zu 
$1/E$ ist. Da die experimentellen Abweichungen nur f"ur gro"se 
Werte von $T/\nu$ auftraten, d.h. bei festem $\nu$ f"ur gro"se 
$E$, modifizierte er diese Proportionalit"at zu $1/E(E+\beta)$ 
mit $\beta=$ konst. Einmalige Integration liefert dann 
sofort das \emph{Planck'sche Strahlungsgesetz}
\begin{equation}
u(T,\nu)=\frac{8\pi\nu^2}{c^3}
\frac{b\nu}{\exp(\frac{a\nu}{T})-1}\,,
\label{Planck}
\end{equation}
welches Planck zur experimentellen Pr"ufung empfahl. Es zeigte 
sich sehr schnell eine gl"anzende "Ubereinstimmung der neuen Formel 
mit den experimentellen Daten. F"ur kleine $T/\nu$ geht (\ref{Planck})
tats"achlich in (\ref{Wien}) "uber, w"ahrend sich im dem Bereich
gro"ser $T/\nu$, in dem sich die experimentellen Abweichungen zeigten,
das "`klassische"' Gesetz (\ref{Rayleigh-Jeans}) ergibt.
Somit ist es als besondere Ironie dieser Geschichte zu verzeichnen, 
dass die Quantentheorie aus Beobachtungen \emph{klassischer} 
Abweichungen von einem nur a posteriori mit Hilfe der 
Quantentheorie zu verstehenden Grenzgesetz entstanden ist.     

\subsection*{Intermezzo: Einsteins Bestimmung der Avogadro-Zahl}
Zu Beginn seiner ber"uhmten Arbeit "uber Lichtquanten aus dem
Jahre 1905 (\cite{Einstein-CW}, Vol.2, Doc.14) macht Einstein eine
wichtige Bemerkung, die man etwa so zusammenfassen kann:
Fordert man, dass (\ref{Rayleigh-Jeans}), was eine notwendige 
Folge der klassischen Physik ist, als Grenzgesetz in der als 
ph"anomenologisch g"ultig angesehenen Planck'schen Formel enthalten
ist, so ergibt sich eine von jeder \emph{theoretischen Begr"undung}
der Planck'schen Formel \emph{unabh"angige} Methode zur 
Bestimmung der Avogadro-Zahl $N$.  Diese Grenzbeziehung gilt 
n"amlich nur, falls 
\begin{equation}
N=\frac{a}{b}\,R\,.
\label{Einstein}
\end{equation}
Da $R$ gut bekannt ist, liefert jede Bestimmung von $a,b$ durch 
Strahlungsmessungen auch einen Wert f"ur $N$. Einstein erhielt 
so den Wert $N=6,17\cdot 10^{23}$, der genau dem zuvor von 
Planck selbst erhaltenen entspricht, allerdings mit einer
von seiner umstrittenen Theorie wesentlich abh"angigen Begr"undung. 
Zu dieser Zeit war dies der mit Abstand genaueste Wert der
Avogadro-Zahl (vgl. Kapitel~5 in \cite{Pais}).
Dabei sei gleich erg"anzt, dass Planck aus der Kenntnis der 
Faradaykonstante (elektrische Ladung eines Mols einwertiger Ionen), 
die aus Elektrolysedaten gut bekannt war, durch Division mit $N$ 
den Wert der elektrischen Elementarladung $e$ erheblich besser 
als je zuvor bestimmte. Wieder ist es als besondere Ironie zu 
verzeichnen, dass dies ausgerechnet durch den damaligen
Anti-Atomisten Planck m"oglich wurde.

\subsection*{Der "`Akt der Verzweiflung"'}
Wie sollte nun Planck nach all seinen M"uhen, das Wien'sche Gesetz 
theoretisch zu zementieren, eine Ableitung des neuen Gesetzes
(\ref{Planck}) herzaubern? Wieder blieb er seiner alten Strategie treu,
(\ref{Planck-Fundamentalgleichung}) zu benutzen und $E(T,\nu)$
"uber $S(E,\nu)$ zu bestimmen. Um letzteres zu erreichen war 
ihm jedes Mittel recht, denn \emph{"`eine theoretische Deutung 
musste um jeden Preis gefunden werden, und w"are er noch so 
hoch"'} \cite{Planck-Brief-1931}. So verfiel er schlie"slich
der bis dahin von ihm bek"ampften statistischen Auffassung der
Entropie durch L.~Boltzmann, wonach die Entropie ${\cal S}$ 
eines Makrozustandes proportional zum Logarithmus der 
Anzahl $W$ seiner mikroskopischen Realisierungen ist. F"ur Gase 
ergibt sich "Ubereinstimmung mit der thermodynamischen 
Definition, wenn als Proportionalit"atsfaktor $k:=R/N$  
gew"ahlt wird, also   
\begin{equation}
{\cal S}=\frac{R}{N}\ln W\,.
\label{Boltzmann}
\end{equation}
In seiner Arbeit vom \emph{14. Dezember 1900} (\cite{Planck-GW},
Bd.1,p.698), in der Planck zum ersten Male eine theoretische 
Begr"undung seiner Gleichung versuchte, wandte er diese Formel auf 
eine Anzahl $n$ gleicher Oszillatoren 
der Frequenz $\nu$ und Gesamtenergie $E_n$ an, wobei $W$ die Anzahl 
der Verteilungen der festen Gesamtenergie auf die $n$ Oszillatoren 
ist. Damit diese Zahl endlich ist, darf $E_n$ nicht beliebig teilbar 
sein. Planck nahm deshalb an, dass die Oszillatoren Energie nur in 
ganzzahligen Vielfachen einer Grundeinheit $\varepsilon$ aufnehmen 
und abgeben k"onnen. \emph{"`Das war eine rein formale Annahme, und 
ich dachte mir nicht viel dabei, sondern eben nur das, dass ich unter 
allen Umst"anden, koste es was es wolle, ein positives Resultat 
herbeif"uhren musste."'} \cite{Planck-Brief-1931}. Man beachte, da"s
Planck an dieser Stelle \emph{nicht} explizit annahm, dass die Energien
der Oszillatoren selbst nur ganzzahlige Vielfache von $\varepsilon$ sein
k"onnen; das hat sp"ater erst Einstein getan. Vielmehr erkl"art Planck 
gleich nach Einf"uhrung des Energiequantums $\varepsilon$ "uber den
Quotienten $p:=E_n/\varepsilon$: \emph{"`Wenn der so berechnete Quotient
keine ganze Zahl ist, so nehme man f"ur $p$ eine in der N"ahe gelegene
ganze Zahl"'}. Setzt man nun $W$ gleich der
Anzahl der M"oglichkeiten, $p$ ununterscheidbare Energiequanten auf $n$ 
(unterscheidbare) Oszillatoren zu verteilen, so erh"alt man 
\begin{equation}
W=\frac{(n+p-1)!}{(n-1)!\cdot p!}\,.
\label{W}
\end{equation}
Eingesetzt in (\ref{Boltzmann}) ergibt
dies\footnote{Man nimmt dazu $n$ und $p$ als gro"se Zahlen an und
verwendet jeweils die Stirling'sche Approximation 
$\ln z!\approx z\ln(z)-z$.}
f"ur die Entropie $S={\cal S}/n$ des einzelnen Oszillators als 
Funktion seiner Energie $E=E_n/n$ (d.h. $p/n=E/\varepsilon$): 
\begin{equation}
S=k[(1+E/\varepsilon)\ln(1+E/\varepsilon)
     -(E/\varepsilon)\ln(E/\varepsilon)]\,.
\label{Entropie}
\end{equation}
Zun"achst sieht man, dass $\varepsilon$ nicht Null sein kann.  
Das Wien'sche Verschiebungsgesetz in der Form (\ref{Wien-Entropie}) 
impliziert weiter, dass $\varepsilon$ proportional $\nu$ sein muss; 
Planck nennt diese Proportionalit"atskonstante $h$. Damit wird 
$\varepsilon=h\nu$ und die Oszillatorenergie $E$ ein ganzzahliges 
Vielfaches davon. Aus (\ref{Entropie}) folgt in nun bekannter 
Weise das Planck'sche Strahlungsgesetz (\ref{Planck}) mit $b=h$ 
und $a=h/k$.

\subsection*{Interpretationen}
"Uber die Bedeutung der (von Planck eingef"uhrten) "`Boltzmann'schen
Konstanten"' $k$ haben wir schon gesprochen; sie ergibt sich aus 
$k=R/N$ und hat mit dem Atomismus zu tun, der $N<\infty$ bedingt.
Die Bedeutung der Planck'schen Konstanten $h$ blieb vorerst dunkel.
Wie war die Quantisierungsbedingung, bei der sich Planck "`nicht viel
dachte"', letztlich zu interpretieren? Als fundamentale Eigenschaft 
mechanisch gedachter Oszillatoren, also im Widerspruch zur klassischen 
Mechanik, oder als fundamentale Eigenschaft der zur Verteilung 
anstehenden Strahlungsenergie, dann w"are sie im Widerspruch zur 
Maxwellschen Theorie. In beiden F"allen w"are der Planck'schen 
Schl"usselgleichung (\ref{Planck-Fundamentalgleichung}) das
Fundament entzogen, wie Einstein in den Jahren ab 1905 wiederholt 
betonte. 

Planck entschied sich f"ur keine dieser M"oglichkeiten, sondern suchte 
die Quantisierung als Folge einer noch unverstandenen Modifikation der 
\emph{Wechselwirkung} zwischen Strahlung und Materie aufzufassen. 
Einen ersten Hinweis darauf, da"s Planck nicht die Quantisierung der 
Oszillatorenenergie selbst meint haben wir bereits oben angegeben.
Aber schon gar nicht wollte sich Planck auf eine Aufgabe der
gerade erst etablierten Maxwellgleichungen einlassen, was er 
nach Aufstellung der in seinen Augen viel zu radikalen
\emph{Lichtquantenhypothese} von 1905 duch A.~Einstein, die 
genau diese Konsequenz heraufbeschwor, auch mehrfach betonte.
So z.B. nach Einsteins viel beachtetem Vortrag "`"Uber die
Entwicklung unserer Anschauungen "uber das Wesen und die 
Konstitution der Strahlung"' anl"a"slich der Tagung der 
Gesellschaft Deutscher Naturforscher und "Arzte 1909 in 
Salzburg (\cite{Einstein-CW}, Vol.2, Doc.60).
Nachdem Einstein eindrucksvoll die Unzul"anglichkeiten der
Planck'schen Strahlungstheorie analysiert und dabei seine 
Lichtquantenhypothese empfiehlt, bemerkt Planck in der 
darauffolgenden Diskussion: \emph{"`Jedenfalls meine 
ich, man m"usste zun"achst versuchen, die ganze Schwierigkeit der
Quantentheorie zu verlegen in das Gebiet der \emph{Wechselwirkung}
zwischen der Materie und der strahlenden Energie; die Vorg"ange im
reinen Vakuum k"onnte man dann vorl"aufig noch mit den Maxwell'schen 
Gleichungen erkl"aren"'} (\cite{Einstein-CW}, Vol2., Doc.61).
In seiner Ablehnung der Einstein'schen Lichtquantenhypothese wurde
Planck "ubigens von fast allen zeitgen"ossischen Physikern unterst"utzt.
Noch 1913 schrieben Planck, Nernst, Rubens und Warburg in ihrem
Empfehlungsschreiben f"ur Einsteins Aufnahme in die Preu"sische
Akademie der Wissenschaften:
\emph{"`Dass er in seinen Spekulationen gelegentlich auch 
einmal  "uber das Ziel hinausgeschossen haben mag, wie z.B. 
in seiner Hypothese der Lichtquanten, wird man ihm nicht 
allzuschwer anrechnen d"urfen"'} (\cite{Einstein-CW}, Vol.5, Doc.445).

So ist auch Plancks Verst"andnis der Abz"ahlung (\ref{W}) nicht das  
einer irgendwie neuen, nicht klassischen Statistik ununterscheidbarer
Objekte (Lichtquanten), so wie sie heute unter dem Namen 
\emph{Bose-Einstein-Statistik} f"ur Photonen aufgefasst wird, sondern 
\emph{rein klassisch}. Genauso wie wenn man nach der Anzahl der
M"oglichkeiten fragt, $p$ Liter Wasser mit einer Sch"opfkelle die 
einen Liter fasst (und immer ganz gef"ullt werden muss) auf $n$ 
Gef"a"se zu verteilen; auch dann w"are nat"urlich (\ref{W}) die 
richtige Antwort. Dem Zwang hier die Sch"opfkelle zur Verteilung 
zu benutzen entspricht bei Planck die noch unverstandene Wechselwirkung 
zwischen Strahlung und Materie, die ebenfalls die Oszillatoren 
zwingt mit dem Kontinuum der Strahlungsenergie immer nur ganze 
Portionen $\varepsilon$ auszutauschen. 

Tats"achlich verfolgte Planck in den Jahren 1911-14 eine
Ab"anderung seiner urspr"unglichen Theorie, in der eine 
Quantisierung des Energieaustauschs nur noch f"ur den
Emissionsvorgang, nicht jedoch f"ur den Vorgang der Absorption 
angenommen werden musste (\cite{Planck-GW}, Bd.2). Als Konsequenz 
ergab sich eine leichte Modifikation des Strahlungsgesetzes 
(\ref{Planck}) um den additiven Term $h\nu/2$, der einer nicht 
verschwindenden Energie des Oszillators beim absoluten Temperaturnullpunkt 
entspricht. Dies ist das erste Auftreten der heute  aus der Quantenmechanik 
wohlbekannten \emph{Nullpunktsenergie}. Erneut bewegt sich Planck 
damit, ohne es zu wissen und diametral gegen seine Intention, von 
der klassischen Theorie weg. Sp"atestens zu diesem Zeitpunkt hatte die 
Entwicklung Planck hinter sich gelassen. Wie kein anderer f"orderte 
Einstein in dieser Zeit durch hartn"ackiges Hinterfragen der 
Grundlagen der Planck'schen Strahlungstheorie den endg"ultigen 
Bruch mit der klassischen Physik; doch das ist eine andere -- 
nicht minder spannende -- Geschichte.  \emph{"`Kurz zusammengefasst"'},
schreibt Planck in dem schon mehrfach zitierten Brief des Jahres 
1931 r"uckschauend, \emph{"`kann ich die ganze Tat als einen Akt der 
Verzweiflung bezeichnen"'} \cite{Planck-Brief-1931}.

\end{document}